\documentclass[aps,prl,reprint,superscriptaddress]{revtex4-2}

\usepackage{graphicx}
\usepackage{dcolumn}
\usepackage{bm}
\usepackage{placeins}

\usepackage{pdfpages}
\usepackage{pgffor}
\makeatletter
\AtBeginDocument{\let\LS@rot\@undefined}
\makeatother

\begin{document}

\title{Magnetic-field-induced superconductivity in hexalayer rhombohedral graphene}

\newcommand{\cofirst}{\altaffiliation{These authors contributed equally.}}

\author{Jinghao Deng}
\affiliation{Laboratory of Atomic and Solid State Physics, Cornell University, Ithaca, NY, USA}

\author{Jiabin Xie}
\affiliation{Laboratory of Atomic and Solid State Physics, Cornell University, Ithaca, NY, USA}

\author{Hongyuan Li}
\affiliation{Laboratory of Atomic and Solid State Physics, Cornell University, Ithaca, NY, USA}

\author{Takashi Taniguchi}
\affiliation{Research Center for Materials Nanoarchitectonics, National Institute for Materials Science,
1-1 Namiki, Tsukuba 305-0044, Japan}

\author{Kenji Watanabe}
\affiliation{Research Center for Electronic and Optical Materials, National Institute for Materials Science,
1-1 Namiki, Tsukuba 305-0044, Japan}

\author{Jie Shan}
\affiliation{School of Applied and Engineering Physics, Cornell University, Ithaca, NY, USA}
\affiliation{Max Planck Institute for the Structure and Dynamics of Matter, Hamburg, Germany}

\author{Kin Fai Mak}
\affiliation{Laboratory of Atomic and Solid State Physics, Cornell University, Ithaca, NY, USA}
\affiliation{Max Planck Institute for the Structure and Dynamics of Matter, Hamburg, Germany}

\author{Xiaomeng Liu}
\email{xl956@cornell.edu}
\affiliation{Laboratory of Atomic and Solid State Physics, Cornell University, Ithaca, NY, USA}

\begin{abstract}
In conventional superconductors, superconductivity is generally suppressed by external magnetic fields due to spin-singlet pairing. Here, we report signatures of in-plane-magnetic-field-induced superconductivity in hexalayer rhombohedral graphene and reveal electric-field control of its depairing behavior. With the application of a small in-plane magnetic field $B_{\parallel}$, a superconducting state emerges within a narrow band along a phase boundary. Its properties evolve continuously with increasing $B_{\parallel}$: the superconducting region progressively shifts toward higher electric field as the $B_{\parallel}$ increases and the transition temperature rises with increasing $B_{\parallel}$. Remarkably, the superconducting state remains robust under $B_{\parallel}$ up to 14 T, far exceeding the conventional Pauli limit. Quantum oscillation measurements further reveal that the superconductivity emerges from nematic Fermi surface reconstruction. These results suggest a spin-polarized superconducting states with unconventional origins.
\end{abstract}

\maketitle

Spin-polarized superconductivity has drawn significant interest due to its connection with spin-triplet pairing and unconventional order parameter~\cite{Uji2001Magneticfieldinducedsuperconductivitytwodimensional, Ran2019Nearlyferromagneticspintriplet, Jiao2020Chiralsuperconductivityheavyfermion}. In contrast to conventional Cooper pairs, spin-triplet superconductor bind electrons with identical spins and corresponds to an odd-parity order parameter, which may break time-reversal symmetry depending on its detailed structure. It may host exotic properties, such as Majorana zero modes, and could be relevant for topological quantum computing~\cite{Sato2017Topologicalsuperconductorsreview}. A common way to identify spin-polarized superconductivity is its resilience~\cite{Ran2019Nearlyferromagneticspintriplet}, or sometimes even reliance on magnetic fields~\cite{Shoenberg1984MagneticOscillationsMetals, Konoike2004Magneticfieldinducedsuperconductivityantiferromagnetic, Ran2019Extrememagneticfieldboosted}. Unlike conventional spin-singlet pairing, which is suppressed by the Zeeman effect of an external magnetic field, called the Pauli limit, spin-polarized superconductors do not incur a Zeeman energy penalty and can therefore remain stable under strong magnetic fields.

Rhombohedral multilayer graphene provides a clean platform for realizing and studying spin-polarized superconductivity with high reproducibility and tunability. Indications of spin-polarized superconductivity were first observed in a trilayer graphene system where the in-plane magnetic critical field is far beyond the Pauli limit~\cite{Zhou2021Superconductivityrhombohedraltrilayer}. Proximity-induced spin-orbit coupling and in-plane magnetic field can further stabilize and even promote the superconductivity in bilayer~\cite{Zhou2022Isospinmagnetismspinpolarized, Li2024Tunablesuperconductivityelectron, Zhang2025Twistprogrammablesuperconductivityspin} and trilayer rhombohedral graphene~\cite{Patterson2025Superconductivityspincanting, Yang2025Impactspinorbita}. Pauli-limit violation behaviors in superconductivity in different parameter spaces are observed in thicker layer systems~\cite{Han2025Signatureschiralsuperconductivity, Deng2025SuperconductivityFerroelectricOrbital}, providing opportunities to investigate spin-triplet pairing and its interplay with different degrees of freedom, including isospin ordering, intervalley coherence, and Fermi surface reconstruction.

In this work, we report transport signatures of spin-polarized superconductivity induced by in-plane magnetic field in a hexalayer rhombohedral graphene system. The superconductivity can persist across a large range of electric and in-plane magnetic fields. We further uncover a close correlation between the electric field and magnetic-field range over which superconductivity emerges, which we attribute to orbital depairings effect arising from electric-field-controlled layer polarization. Quantum oscillation measurements identify the parent phase of the spin-polarized superconductivity as a nematic, partially isospin-polarized state. These results establish hexalayer rhombohedral graphene as a promising platform for exploring spin-polarized superconductivity.

Fig. 1(a) and (b) show the longitudinal resistance $R_{xx}$ measured at in-plane magnetic field $B_{\parallel}$ = 0 and 0.5 T at the base temperature of a dilution fridge (see Supplemental Materials~\cite{SM} for experimental methods). Notably, we observe a butterfly-shaped phase boundary, marked by the dashed line in Fig. 1a, across which the resistance exhibits a sudden change (see Supplemental Materials~\cite{SM} for calculations of $n$ and $E$). This boundary likely indicates a Lifshitz transition---a change of the Fermi surface topology. Surprisingly, the application of small in-plane magnetic field $B_{\parallel}$ = 0.5 T led to a resistance dip along low electric field regime of this phase boundary, suggesting a possible superconducting phase transition triggered by the in-plane magnetic field (Fig. 1(b)). 

\FloatBarrier
\begin{figure*}[t]
  \centering
  \includegraphics[width=\textwidth]{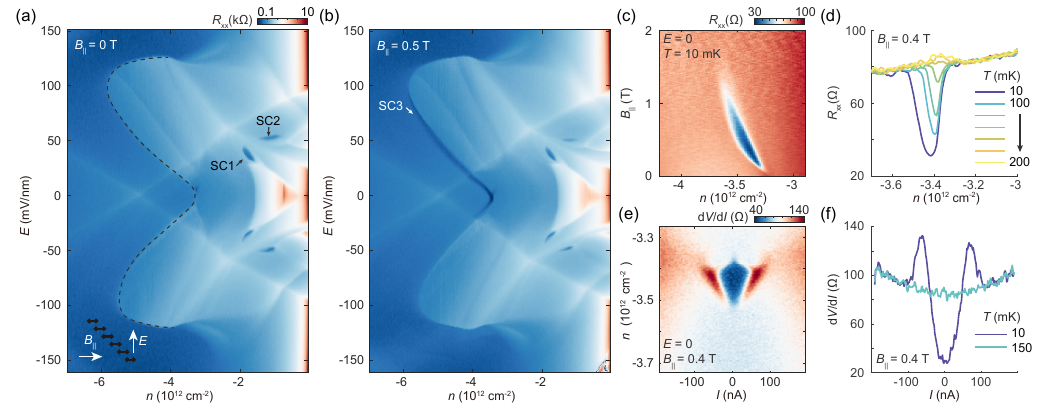}
  \caption{In-plane-field-induced superconductivity in rhombohedral hexalayer graphene.
  (a),(b) Color-coded $R_{xx}$ as a function of carrier density $n$ and electric field $E$ at zero field and at an in-plane magnetic field $B_{\parallel}=0.5~\mathrm{T}$.
  The dashed curves indicate the phase boundary from which superconductivity arises.
  SC1 and SC2 correspond to previously reported superconducting states in this system~\cite{Deng2025SuperconductivityFerroelectricOrbital}.
  SC3 denotes the superconducting phase observed in this work. The inset in panel~(a) illustrates rhomhehdral hexalayer graphene under electric field and inplane magnetic field.
  (c) $B_{\parallel}$-dependent $R_{xx}$--$n$ line cuts at $E=0$. At this specific electric field, superconductivity arises with $B_{\parallel} \approx 0.07~\mathrm{T}$ and disappears above $B_{\parallel} \approx 1.2~\mathrm{T}$
  (d) Temperature dependence of $R_{xx}$ vs $n$ at $B_{\parallel}=0.4~\mathrm{T}$.
  (e) Differential resistance $dV/dI$ at $E=0$ and $B_{\parallel}=0.4~\mathrm{T}$.
  (f) $dV/dI$ at $E=0$, $B_{\parallel}=0.4~\mathrm{T}$, and
  $n=-3.46 \times 10^{12}~\mathrm{cm}^{-2}$, measured below and above the transition temperature at $10~\mathrm{mK}$ and $150~\mathrm{mK}$, respectively.}
  \label{fig:1}
\end{figure*}

The in-plane-field-induced superconducting-like state clearly manifest in Fig.~1c. Line cuts of $R_{xx}$ versus $B_{\parallel}$ at zero electric field reveal a resistance dip that emerges above $B_{\parallel}\approx0.07~\mathrm{T}$ and disappears beyond $B_{\parallel}\approx1.2~\mathrm{T}$. This in-plane-field response closely resembles the spin-polarized superconductivity previously reported in Bernal bilayer graphene~\cite{Zhou2022Isospinmagnetismspinpolarized}.

With increasing $B_{\parallel}$, the superconducting dip shifts toward higher hole densities (Fig.~1(c)), forming an arc, likely due to a Zeeman-effect-induced shift of the phase boundary between a spin-unpolarized metal on the left and a spin-polarized metal on the right. The superconducting feature appears closely connected to the Lifshitz transition: in Fig.~1(c), extrapolation of the superconducting arc toward zero $B_{\parallel}$ intersects the step-like feature in the normal-state resistance.

Further evidence for this connection is provided by the temperature dependence at finite $B_{\parallel}$. Fig.~1(d) shows density-dependent $R_{xx}$ measured at $B_{\parallel}=0.4~\mathrm{T}$ and $E=0$. As temperature increases, the resistance dip disappears, while the resistance peak associated with the transition between spin-polarized and spin-unpolarized states reappears on the higher-density side of the superconducting region. This behavior supports a direct connection between superconductivity and the Lifshitz transition above the critical temperature. The superconducting dip consistently occurs on the spin-unpolarized side of the phase boundary, suggesting that the in-plane magnetic field polarizes a nearby unpolarized state and stabilizes a superconducting ground state.

The superconducting character is further tested through $I$–$V$ measurements. Figures~1(e) and 1(f) present differential resistance $dV/dI$ measured at $B_{\parallel}=0.4~\mathrm{T}$ and $E=0$. Compared with $T=150~\mathrm{mK}$ [Fig.~1(f)], $dV/dI$ at base temperature exhibits pronounced nonlinear characteristics of a superconductor. The two peaks in $dV/dI$, corresponding to the superconducting critical current, reach their maximum near the Lifshitz-transition boundary and are suppressed toward both the spin-unpolarized and weakly polarized regimes.

Taken together, these measurements provide strong evidence for in-plane field induced superconductivity at a Lifshitz transition. We note that a finite residual resistance of approximately $37\%$ of the normal-state value remains. Such residual resistance is commonly observed in rhombohedral graphene systems with low superconducting transition temperatures ~\cite{Holleis2025Nematicityorbitaldepairing, Kumar2025Superconductivitydualsurfacecarriers}.

\FloatBarrier
\begin{figure*}[t]
  \centering
  \includegraphics[width=\textwidth]{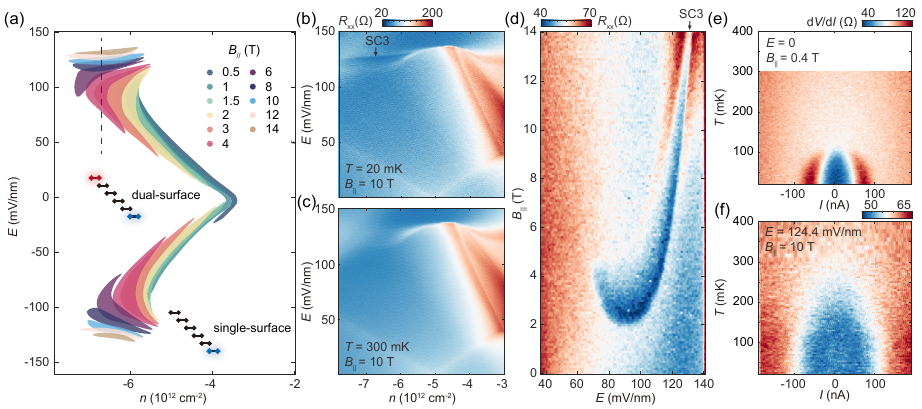}
  \caption{Tunable superconductivity controlled by in-plane magnetic field and electric field.
  (a) Schematic of the superconducting phase space under various in-plane magnetic field, constructed from a series of $n$--$E$--$R_{\mathrm{xx}}$ maps (see Supplemental Material).
  The inset shows schematic wave-function weight distributions in rhombohedral hexalayer graphene at zero and large $E$, where red and blue indicate localization near the top and bottom surfaces, respectively.
  (b,c) $n$--$E$--$R_{\mathrm{xx}}$ maps measured at $(T,~B_\parallel)$ = (20 mK, 10 T) and (300 mK, 10 T), respectively. 
  (d) $R_{xx}$ vs $E$ and $B_{\parallel}$ measured at a fixed carrier density of $n = -6.57 \times 10^{12}~\mathrm{cm}^{-2}$. The position of this linecut in $n$-$E$ space is indicated by a vertical dashed line in (a).
  (e,f) Temperature-dependent differential resistance $dV/dI$ measured at $B_{\parallel} = 0$ and $B_{\parallel} = 10~\mathrm{T}$, respectively. 0 T data is measured at $E = 0$, $n=-3.46 \times 10^{12}~\mathrm{cm}^{-2}$, and $10~\mathrm{T}$ data is measured at $ E = 124.4~\mathrm{mV/nm}$ and $n=-6.57 \times 10^{12}~\mathrm{cm}^{-2}$.
  }
  \label{fig:2}
\end{figure*}

To elucidate the evolution of superconductivity under an in-plane magnetic field, we measured a series of $n$–$E$–$R_{xx}$ maps at various $B_{\parallel}$ up to $14~\mathrm{T}$ and summarize the resulting superconducting phase space in Fig.~2(a) (see Supplemental Material~\cite{SM} for the complete dataset). Representative high-field $R_{xx}$ maps at $B_{\parallel} = 10~\mathrm{T}$ are shown in Figs.~2(b) and (c), taken below and above the superconducting transition temperature, respectively. Notably, increasing magnetic field suppresses superconductivity in the low-$E$ region and causes the superconducting phase to shift toward higher electric fields. Strikingly, in the high $E$ regime, superconductivity persists up to $14~\mathrm{T}$, the maximum field accessible in our experiment [Fig.~2(d)].
To the best of our knowledge, this value represents the highest in-plane critical magnetic field reported to date in rhombohedral multilayer graphene systems.

Figures~2(e) and (f) present representative temperature-dependent $dV/dI$ measurements performed at $(E,B_{\parallel})=(0,0.4~\mathrm{T})$ and $(124.4~\mathrm{mV/nm},10~\mathrm{T})$, respectively. Remarkably, the superconducting state at $B_{\parallel}=10~\mathrm{T}$ exhibits an enhanced critical temperature of $T_{\mathrm{c}}\approx260~\mathrm{mK}$, compared with $T_{\mathrm{c}}\approx110~\mathrm{mK}$ at $B_{\parallel}=0.4~\mathrm{T}$, together with a significantly increased critical current. A complete set of density-, electric-field-, and temperature-dependent $dV/dI$ measurements is provided in the Supplemental Material~\cite{SM}. These results demonstrate that neither strong electric fields nor large in-plane magnetic fields suppress superconductivity; instead, they enhance it.

For conventional BCS superconductors, the Zeeman effect polarizes electron spins, breaking spin-singlet Cooper pairs and suppressing superconductivity at the Pauli-limited $B_{\mathrm{P}}=1.86~\mathrm{T/K}\times T_{\mathrm{c}}$. Using the critical temperature $T_{\mathrm{c}}\approx110~\mathrm{mK}$ measured at $B_{\parallel}=0.4~\mathrm{T}$, the corresponding Pauli limit is approximately $0.2~\mathrm{T}$. At zero electric field, superconductivity survives up to a critical field of approximately $1.2~\mathrm{T}$, already exceeding the Pauli limit. Under large electric fields, superconductivity further persists up to $14~\mathrm{T}$ with an enhanced critical temperature, exceeding the Pauli limit by nearly two orders of magnitude. These observations indicate that the superconducting state is immune to Zeeman pair breaking and provide strong evidence for spin-triplet pairing.

The lower critical field observed at zero electric field can be understood in terms of orbital depairing. Beyond Zeeman pair breaking, disorder, Rashba spin--orbit coupling (SOC), and orbital effects are known to suppress superconductivity~\cite{Tinkham2004Introductionsuperconductivity, Zhang2023Enhancedsuperconductivityspin, Holleis2025Nematicityorbitaldepairing}. Disorder can strongly suppress non-$s$-wave superconductivity, but its effect is not expected to depend sensitively on electric field. The intrinsic SOC in rhombohedral graphene is negligibly small, suggesting that orbital depairing is the dominant mechanism limiting the critical field. In a two-dimensional superconductor with finite thickness, an in-plane magnetic field can penetrate through the sample edges and thread magnetic flux, leading to orbital depairing characterized by a critical field of order
\[
B_{c\parallel}^{\mathrm{orb}} \sim \frac{\Phi_0}{\xi\, d_{\mathrm{eff}}},
\]
where $\Phi_0=h/2e$ is the superconducting flux quantum, $\xi$ is the coherence length, and $d_{\mathrm{eff}}$ is the effective thickness~\cite{Tinkham2004Introductionsuperconductivity}.

In rhombohedral hexalayer graphene, the low-energy electronic states are strongly localized at the top and bottom layers, separated by approximately $1.7~\mathrm{nm}$. At zero electric field, in the absence of correlation-induced polarization, the electronic states are symmetric superpositions of the top and bottom layers, susceptible of orbital depairing. Under large electric fields, however, the electronic wavefunctions become strongly polarized into the top or bottom layer. As a result, the effective thickness $d_{\mathrm{eff}}$ is significantly reduced, providing a natural explanation for the enhanced robustness of superconductivity. This electric-field-tunable orbital depairing suggests that the zero-field superconducting state involves contributions from both surfaces.
\begin{figure}[t]
  \centering
  \includegraphics[width=\linewidth]{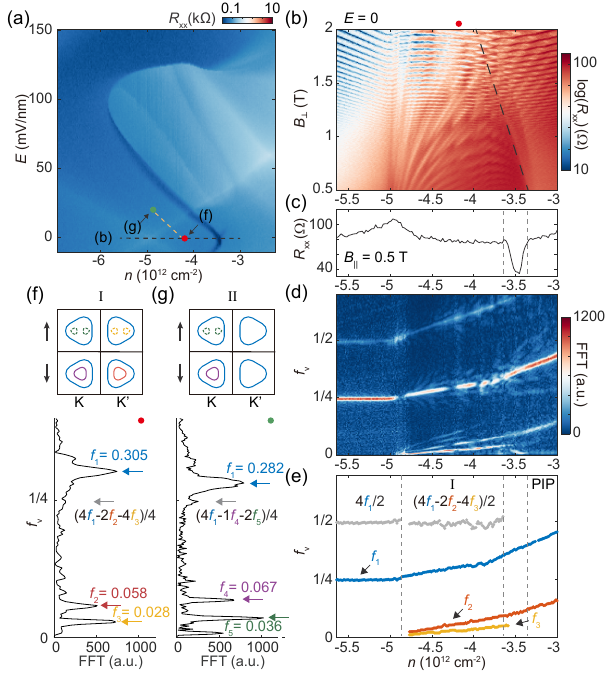}
  \caption{Fermiology around the superconductivity.
  (a) $R_{xx}$–$n$–$E$ map at $B_{\parallel}=0.5~\mathrm{T}$. The black and orange dashed lines indicate the SdH measurement paths corresponding to panels (b) and Fig.~4, respectively. The red and green dots mark the locations of the SdH measurements shown in panels (f) and (g).
(b) $R_{xx}$ as a function of carrier density $n$ and out-of-plane magnetic field $B_{\perp}$.
(c) $R_{xx}$ versus carrier density $n$ at $B_{\perp}=0$ and in-plane magnetic field $B_{\parallel}=0.5~\mathrm{T}$, showing density range of superconductivity.
(d), (e) FFT analysis of the quantum oscillations and the extracted peak frequencies at $E=0$. Distinct peaks $f_{\nu}$ are indicated by colored markers in panel (e). The corresponding sum-rule results are shown as gray markers and are divided by 2 for clarity.
(f), (g) Top panels show schematic Fermi-surface contours for representative phases corresponding to the red and green markers in panel (a). The labels I and II correspond to the same labels in panel (e) and Fig.~4(b). In each schematic, four pockets represent an approximate fourfold degeneracy arising from spin ($\uparrow,\downarrow$) and valley ($K,K'$) degrees of freedom. 
Bottom panels show the FFT analysis of the quantum oscillations and the corresponding sum-rule calculations for the $(n,E)$ positions indicated by the red and green dots in panel (a).}
  \label{fig:3}
\end{figure}

We next carry out a fermiology analysis, which indicates that superconductivity emerges from an isospin-polarized nematic state. Figure~3(b) shows the quantum oscillations under an out-of-plane magnetic field $B_{\perp}$ at $E=0$. A marked change in the oscillation pattern occurs across a phase boundary indicated by the dashed line. This boundary shifts with increasing magnetic field, similar to the phase-boundary shift observed under an in-plane magnetic field in Fig.~1(c). The superconducting states notably reside on the high hole-doping side of this phase boundary (Fig.~3c).

We perform fast Fourier transforms (FFT) of the quantum oscillations and normalize the resulting frequencies by the carrier density, yielding the normalized quantum oscillation frequency 
$f_{\nu} = \frac{e}{nh}\,f_{\mathrm{FFT}}$, 
where $f_{\mathrm{FFT}}$ is the oscillation frequency of $R_{xx}$ as a function of $1/B_{\perp}$, and $e$ and $h$ are the electron charge and Planck constant, respectively~\cite{Shoenberg1984MagneticOscillationsMetals}. This normalized frequency, plotted in Fig.~3(d), represents the fraction of the Fermi-surface area associated with each peak relative to the total occupied Fermi-surface area.

On the high hole-doping side of the phase boundary, where superconductivity emerges, three prominent frequency peaks $f_1$, $f_2$, and $f_3$ can be identified, marked by blue, red, and orange line in Figs.~3(e). The sum of these frequencies should satisfy the sum rule $\sum s k f_{\nu} = 1$, where $s=\pm1$ corresponds to hole-like and electron-like Fermi surfaces in the hole-doped regime, and $k$ represents the degeneracy of the Fermi surface. For a spin- and valley-degenerate metallic state with an annular Fermi surface, we expect $4f_{\mathrm{major}} - 4f_{\mathrm{minor}} = 1$, where $f_{\mathrm{major}}$ and $f_{\mathrm{minor}}$ denote the outer and inner Fermi contours, respectively. Here we instead observe an unexpected sum rule $4f_{1} - 2f_{2} - 4f_{3} = 1$ [Figs.~3(e) and 3(f)].

One possible Fermi-surface structure consistent with this sum rule is illustrated in Fig.~3(f). For two isospin flavors, the inner annular Fermi contour splits into two smaller contours that contribute to the frequency $f_3$. The unsplit inner annular contour of the remaining two flavors corresponds to $f_2$, together with a fourfold-degenerate outer Fermi contour corresponding to $f_1$. Such splitting can be phenomenologically attributed to trigonal-warping effects near the band edge of rhombohedral multilayer graphene, in conjunction with nematicity that breaks the $C_{3}$ crystalline rotational symmetry~\cite{Huang2023Spinorbitalmetallic, Koshino2009TrigonalwarpingBerrys, Holleis2025Nematicityorbitaldepairing}.

At the low hole-doping side of the phase boundary, the FFT spectra exhibit two major frequency peaks, but no sum rule can be found. Such behavior is often found in iso-spin imbalanced states, where hole occupation in one spin band is greater than the other, also known as partial-isospin-polarized (PIP) state~\cite{Zhou2021Halfquartermetalsrhombohedral}. This state features a net spin polarization, which is consistent with shifting of the phase boundary under magnetic fields shown in Fig.~1(c) and 3(b).

\begin{figure}[t]
  \centering
  \includegraphics[width=\linewidth]{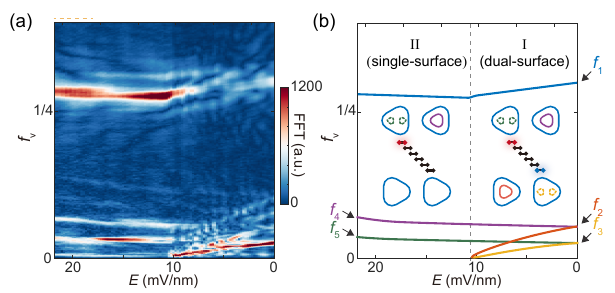}
  \caption{Dual-surface to single-surface nematic Fermi surface transition.
(a) Linecut of carrier-density-normalized FFT from $R_{xx}$ as a function of $1/B_{\perp}$ along the orange dashed line in Fig. 3(a). 
(b) Sketch of the frequency peaks in (a). Inset shows schematic  diagram of dual-surface and single-surface nematicity behavior in regions I and II, respectively}
  \label{fig:4}
\end{figure}

At $|E| > 10~\mathrm{mV/nm}$, we observe a different sum rule: $4f_{1}-f_{4}-2f_{5}=1$,
as shown in Fig.~3(g) for measurements taken at the green dot in Fig.~3(a). 
Compared to the $E=0$ case, the degeneracy of both the inner annular Fermi surface and the nematic Fermi pockets is reduced by a factor of two, suggesting the Fermi-surface configuration illustrated in Fig.~3(g).

The reduction of degeneracy is further captured in Fig.~4, which shows quantum oscillations measured along the path marked by the orange dashed line in Fig.~3(a). Under electric field, we observe that $f_{2}$ and $f_{3}$ split into two distinct branches: $f_2$ and $f_3$; $f_4$ and $f_5$. As the electric field increases, $f_2$ and $f_3$ gradually decrease and eventually disappear for $E>10~\mathrm{mV/nm}$. This contrasting electric-field response suggests that the two branches originate from Fermi surfaces localized on the opposite graphene layers. At $E=0$, electron states are evenly distributed between the top and bottom graphene layer. As the electric field increases, electrons become increasingly polarized toward one layer, while $f_2$ and $f_3$ associated with the opposite layer shrink and ultimately vanish, leaving only a single branch of the Fermi surfaces, $f_4$ and$f_5$, for $E>10~\mathrm{mV/nm}$. This transition from a dual-surface nematic state to a single-surface nematic state is consistent with the observed electric-field modulation of orbital depairing. 

In summary, our study reveals an in-plane-field-induced superconducting phase that survives under remarkably large in-plane magnetic fields. This suggests a spin-polarized superconducting state. The enhanced critical in-plane field at larger electric fields indicates a reduced effective thickness of the superconducting state due to layer polarization, which suppresses orbital depairing. Consistently, our fermiology analysis supports this layer-polarization transition and identifies nematic Fermi surface structures from which superconductivity emerges.



\emph{Note added.}—
During the preparation of this work, we became aware of several related studies reporting in-plane-magnetic-field-induced or enhanced superconductivity~\cite{Kumar2025Pervasivespintripletsuperconductivity, Guo2025Flatbandsurface, Yang2025MagneticFieldEnhancedGraphene, Xie2025MagneticFieldDrivenInsulatorSuperconductorTransition}.

\section*{Acknowledgments}
X. L. acknowledges support from the National Science Foundation through the CAREER program under Award No.~DMR-2442363.
J. D. acknowledges support from the New Frontier Grant, College of Arts \& Sciences, Cornell University.
The device fabrication is performed in part in Cornell Center for Materials Research and in part at the Cornell NanoScale Facility, a member of the National Nanotechnology Coordinated Infrastructure (NNCI), which is supported by the National Science Foundation (Grant NNCI-2025233).
K.W. and T.T. acknowledge support from the JSPS KAKENHI (Grant Numbers 21H05233 and 23H02052), the CREST (JPMJCR24A5), JST and World Premier International Research Center Initiative (WPI), MEXT, Japan.

\bibliographystyle{apsrev4-2}
\bibliography{Bib_SPSC}

\clearpage
\onecolumngrid

\foreach \x in {1,...,4}{
\begin{figure*}[p]
\centering
\includegraphics[width=\textwidth,page=\x]{SI_SP_SC_R6L_v1.pdf}
\end{figure*}
\clearpage
}

\end{document}